%% file: main.tex
\documentclass[journal]{IEEEtran}

\usepackage{amsmath,amsfonts}
\usepackage{algorithmic}
\usepackage{algorithm}
\usepackage{array}
\usepackage[caption=false,font=normalsize,labelfont=sf,textfont=sf]{subfig}
\usepackage{textcomp}
\usepackage{stfloats}
\usepackage{url}
\usepackage{verbatim}
\usepackage{graphicx}
\usepackage{cite}
\usepackage{xr-hyper}
\usepackage[hidelinks]{hyperref}

\usepackage{gensymb} % For degree
\usepackage{booktabs}

\newcommand{\DC}{H}
\newcommand{\MM}{G}
\newcommand{\DM}{F}
\newcommand{\FR}{R}
\newcommand{\posit}{r}
\newcommand{\kposit}{k}
\newcommand{\fsdit}{m}
\newcommand{\displit}{u}
\newcommand{\velit}{v}
\newcommand{\gencoorit}{q}
\newcommand{\bfcnit}{\phi}
\newcommand{\bfcnmat}{\Phi}

\newcommand{\forceit}{f}
\newcommand{\measd}{\mathbf{d}}

\newcommand{\encop}{E}
\newcommand{\fourierop}{F}
\newcommand{\Nfull}{N}
\newcommand{\Nmeas}{M}
\newcommand{\Ntime}{T}
\newcommand{\Ndim}{d}
\newcommand{\idim}{j}
\newcommand{\NDOF}{P}
\newcommand{\iDOF}{p}
\newcommand{\Ndynpar}{1}
\newcommand{\lDM}{\lambda_F}
\newcommand{\lDC}{\lambda_H}
\newcommand{\lFR}{\lambda_R}
\newcommand{\orient}{\alpha}
\newcommand{\damping}{c}
\newcommand{\stiffness}{\kappa}

\newcommand{\pos}{\mathbf{\posit}}
\newcommand{\kpos}{\mathbf{\kposit}}
\newcommand{\fsd}{\mathbf{\fsdit}}
\newcommand{\displ}{\mathbf{\displit}}
\newcommand{\vel}{\mathbf{\velit}}
\newcommand{\gencoor}{\mathbf{\gencoorit}}
\newcommand{\bfcn}{\boldsymbol{\bfcnit}}
\newcommand{\force}{\mathbf{\forceit}}

\begin{document}

\title{Time-Resolved Reconstruction of Motion, Force, and Stiffness using Spectro-Dynamic MRI}

\author{Max H. C. van Riel, Tristan van Leeuwen, Cornelis A. T. van den Berg, Alessandro Sbrizzi
        % <-this % stops a space
\thanks{This work was supported by a grant by the Nederlandse Organisatie voor Wetenschappelijk Onderzoek (NWO) under grant number 18897. \textit{(Corresponding author: Max H. C. van Riel)}}% <-this % stops a space
\thanks{M. H. C. van Riel, C. A. T. van den Berg and A. Sbrizzi are with the Department of Radiotherapy, Computational Imaging Group for MR Diagnostics and Therapy, UMC Utrecht, Utrecht, The Netherlands (e-mail: m.h.c.vanriel-3@umcutrecht.nl).}%
\thanks{T. van Leeuwen is with the Mathematical Institute, Utrecht University, Utrecht, The Netherlands.}%
% \thanks{This paper has supplementary material available at http://ieeexplore.ieee.org provided by the authors.}%
\thanks{\copyright~2023 IEEE. Personal use of this material is permitted. Permission from IEEE must be obtained for all other uses, in any current or future media, including reprinting/republishing this material for advertising or promotional purposes, creating new collective works, for resale or redistribution to servers or lists, or reuse of any copyrighted component of this work in other works.}
}

\maketitle

\begin{abstract}
\input{0_Abstract}
\end{abstract}


\section{Introduction}
\input{1_Introduction}

\section{Theory}
\input{2_Theory}

\section{Methods}
\input{3_Methods}

\section{Results}
\input{4_Results}

\section{Discussion}
\input{5_Discussion}

\section{Conclusion}
\input{6_Conclusion}

\section*{Acknowledgment}
The authors would like to thank David Heesterbeek for fruitful discussions.

\bibliographystyle{IEEEtran}
\bibliography{references.bib}

\newpage

\input{7_Biography}

\vfill

\newpage
\onecolumn
\input{supplementary}

\end{document}

%% file: 0_Abstract.tex
Measuring the dynamics and mechanical properties of muscles and joints is important to understand the (patho)physiology of muscles. However, acquiring dynamic time-resolved MRI data is challenging. We have previously developed Spectro-Dynamic MRI which allows the characterization of dynamical systems at a high spatial and temporal resolution directly from k-space data. This work presents an extended Spectro-Dynamic MRI framework that reconstructs 1) time-resolved MR images, 2) time-resolved motion fields, 3) dynamical parameters, and 4) an activation force, at a temporal resolution of 11 ms. An iterative algorithm solves a minimization problem containing four terms: a motion model relating the motion to the fully-sampled k-space data, a dynamical model describing the expected type of dynamics, a data consistency term describing the undersampling pattern, and finally a regularization term for the activation force. We acquired MRI data using a dynamic motion phantom programmed to move like an actively driven linear elastic system, from which all dynamic variables could be accurately reconstructed, regardless of the sampling pattern. The proposed method performed better than a two-step approach, where time-resolved images were first reconstructed from the undersampled data without any information about the motion, followed by a motion estimation step.

%% file: 1_Introduction.tex
\IEEEPARstart{D}{ynamic} measurements of joints and muscles are important to study the (patho)physiology of the musculoskeletal system \cite{Shellock2003FunctionalImaging, Gold2003DynamicImaging, Shapiro2012MRIMovement, Borotikar2017DynamicDisorders, Alcazar2019OnDouble-Hyperbolic}. Previous studies have shown that dynamic scans result in different kinematics compared to static scans \cite{dEntremont2013DoMethods}. Mechanical properties of biological tissues can give additional quantitative information, but these values are often obtained ex vivo or (quasi-)statically \cite{Bilston2015MeasurementChallenges, Guimaraes2020TheEngineering}. MRI is a promising imaging modality for measuring these dynamic properties in vivo, as it provides excellent soft-tissue contrast. However, acquiring dynamic MRI data at a high spatial and temporal resolution remains a challenge. 

Several techniques exist for dynamic musculoskeletal imaging using MRI. Some methods use gating to bin the data into different motion states \cite{Sheehan1998UsingDynamics, Kaiser2013MeasurementSampling}. In contrast, real-time MRI does not require periodic motion or synchronization of the acquisition to the motion \cite{Asakawa2003Real-TimeVelocity, Nayak2022Real-TimeImaging}. Sparsity along the temporal dimension can be exploited to reconstruct dynamic musculoskeletal images from undersampled data \cite{Mazzoli2017AcceleratedKinematics,Shaw2019Real-timeInstability,Menon2020InGRASP-MRI}. Real-time methods are preferred for musculoskeletal applications, since accurately repeating the same motion is difficult to achieve, and sometimes not feasible for patients who experience pain. Therefore, the development of time-resolved 3D MRI techniques with a high temporal resolution is an active area of research \cite{Feng2020GRASP-Pro:Automation}.

All methods described so far extract dynamic information from a time series of images. However, to achieve a sufficient temporal resolution, high undersampling factors are required. Any undersampling artifacts remaining in the images will have a detrimental effect on the estimated dynamics. It has been shown that dynamic information can be extracted directly from k-space data, even in the case of very high undersampling factors, without an intermediate image reconstruction step \cite{Huttinga2020Non-rigidMR-MOTUS}.

In our previous work \cite{VanRiel2022Spectro-DynamicScale}, we have proposed the Spectro-Dynamic MRI framework in an effort to characterize dynamical systems at a high temporal resolution. This novel acquisition paradigm uses a spectral motion model, which relates the raw MRI measurements in the spectral domain (k-space) to the displacement field. Thus, it can reconstruct the motion directly from k-space data, without requiring images at a high temporal resolution as an intermediate step. In addition, it uses a dynamical model which adds prior knowledge about the structure of the motion. This dynamical model also allows for the estimation of dynamical parameters from extremely undersampled data. We demonstrated Spectro-Dynamic MRI using two coupled spherical pendula, from which the motion, the length of the pendula, and the spring stiffness could be estimated accurately. 

In this work, we address some of the limitations of the Spectro-Dynamic MRI reconstruction as presented in \cite{VanRiel2022Spectro-DynamicScale}. Firstly, the reconstruction method required rigid motion fields, or at most motion fields that are spatially varying along the readout direction, due to the high undersampling factor. This prevented the reconstruction of more general motion fields. Secondly, the measurement noise was not explicitly taken into account in the reconstruction. This could lead to biases in the estimated motion.

Here, we propose an extended iterative reconstruction framework for Spectro-Dynamic MRI to overcome these limitations. Besides the motion field and dynamical parameters, the full time-resolved k-space is added as an additional variable in the reconstruction. By jointly reconstructing the missing k-space data and the motion fields in a single optimization problem, a very high undersampling factor per time instance could be achieved. In addition, we have added an external activation force, which allows for the dynamical model to display the output of an arbitrary forcing function. This activation represents an external load acting on a muscle, or neuronal activation of a muscle. Using a dynamic motion phantom, we show that we can reconstruct 1) time-resolved images, 2) time-resolved motion fields, 3) dynamical parameters, and 4) the activation force, at a temporal resolution of 11 ms and a spatial resolution of 5.0 mm $\times$ 5.0 mm. These results bring us closer to our goal of real-time in vivo dynamical system characterization.

%% file: 2_Theory.tex
\begin{table}[!t]
    \centering
    \caption{Table of variables.}
    \begin{tabular}{cc}
        \toprule
        \textbf{Symbol} & \textbf{Description} \\
        \midrule
        \textbf{bold} & Vector quantity \\
        $\hat{}$ & Spectral quantity \\
        % $*$ & Convolution \\
        % $i$ & Imaginary unit \\
        $\pos, \kpos$ & Spatial/Spectral coordinate \\
        % $\pos$ & Spatial coordinate \\
        % $\kpos$ & Spectral coordinate \\
        % $\Omega_r, \Omega_k, \Omega_t$ & Spatial/Spectral/Temporal domain \\
        % $\Omega_{rt}, \Omega_{kt}$ & Spatio/Spectro-temporal domain \\
        % $\Omega_r$ & Spatial domain \\
        % $\Omega_k$ & Spectral domain \\
        % $\Omega_t$ & Time interval \\
        % $\Omega_{rt}$ & Spatio-temporal domain \\
        % $\Omega_{kt}$ & Spectro-temporal domain \\
        $\Omega_M, \Omega_S$ & Moving/Stationary compartment \\
        % $\Omega_M$ & Moving compartment \\
        % $\Omega_S$ & Stationary compartment \\
        $\Ndim$ & Number of dimensions \\
        $\NDOF$ & Number of basis functions \\
        $\Nfull$ & Number of k-space samples per time instance \\
        $\Nmeas$ & Number of measured samples per time instance \\
        $\Ntime$ & Number of time instances \\
        $\fsdit, \hat{\fsdit}$ & Time-resolved image/k-space data \\
        % $\hat{\fsdit}$ & Time-resolved k-space data \\
        $\vel, \displ$ & Velocity/Displacement field \\
        % $\displ$ & Displacement field \\
        $\mathcal{F}$ & Fourier transform operator \\
        % $\encop$ & Sampling operator \\
        % $\bfcnmat$ & Basis function operator \\
        % $C(\cdot,\cdot)$ & Convolution operator \\
        % $D_t$ & First-order temporal finite difference operator \\
        % $D_{tt}$ & Second-order temporal finite difference operator \\
        % $I$ & Identity operator \\
        $\measd$ & Undersampled k-space data \\
        $\bfcn$ & Basis function of the displacement field \\
        $\fsd$ & Time-resolved k-space \\
        $\gencoor$ & Motion field coefficients \\
        $\stiffness$ & Elastic stiffness \\
        $\damping$ & Damping coefficient \\
        $\force$ & Activation force \\
        $\MM$ & Motion model \\
        $\DM$ & Dynamical model \\
        $\DC$ & Data consistency model \\
        $\FR$ & Regularization \\
        $\lDM, \lDC, \lFR$ & Regularization parameters \\
        $\orient$ & Readout rotation angle \\
        \bottomrule
    \end{tabular}
    \label{tab:variables}
\end{table}

\begin{figure}[!t]
    \centerline{\includegraphics[width=\columnwidth]{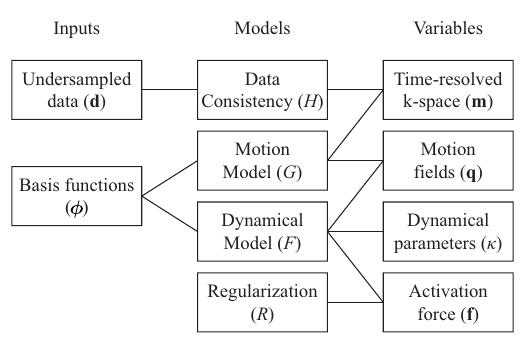}}
    \caption{Overview of the different elements used in the iterative reconstruction algorithm. The four different model components are indicated in the middle column. The measured undersampled data ($\measd$) and the chosen basis functions ($\bfcn$) are the inputs of the reconstruction, and remain constant during the reconstruction. The optimization variables that are reconstructed are given on the right. The connections indicate which inputs and variables are used in each model. Note that every variable is shared between at least two models.}
    \label{fig:iter-algo}
\end{figure}

The extended Spectro-Dynamic MRI reconstruction framework contains four components (Fig. \ref{fig:iter-algo}): a motion model ($\MM$), a dynamical model ($\DM$), a data consistency term ($\DC$), and a regularization term ($\FR$). These four components are combined as penalty terms in a minimization problem. We use penalty terms instead of hard equality constraints to capture model inaccuracies and noise in all four components \cite{vanLeeuwen2016AProblems}. Each component is discussed in more detail in the following subsections. The minimization problem is solved with an iterative reconstruction algorithm, which will be described in Section \ref{subsec:iter-recon}. Some symbols used throughout this work are listed in Table \ref{tab:variables}.

\subsection{Motion Model}
\label{subsec:motion-model}
The motion model $\MM$ provides the relation between the MRI data and the motion field. Its derivation starts by assuming that all signal is conserved during motion \cite{VanRiel2022Spectro-DynamicScale}. This assumption holds if the magnetization is in steady state, the receive and excitation fields are homogeneous, and the readouts are short. 

Let $\Omega_r \subseteq \mathbb{R}^\Ndim$ be a spatial domain in $\Ndim$ dimensions, and let $\Omega_t \subseteq \mathbb{R}$ be a time interval. Then $\Omega_{rt} = \Omega_r \times \Omega_t$ is the combined spatio-temporal domain. In this work, only 2D experiments are shown, but all methods are valid in 3D as well. The extension to 3D will be the topic of future research. 

Let $\fsdit(\pos,t) : \Omega_{rt} \to \mathbb{C}$ be a temporal series of complex-valued images, and let $\vel(\pos,t) : \Omega_{rt} \to \mathbb{R}^\Ndim$ be the $\Ndim$-dimensional velocity field. The conservation of magnetization results in the following partial differential equation (PDE), known as the continuity equation:

\begin{equation}
    \label{eq:continuity-equation}
    \frac{\partial \fsdit}{\partial t} + \nabla \cdot \left(\fsdit \vel \right) = 0.
\end{equation}

Note that combined with the incompressibility constraint ($\nabla \cdot \vel = 0$), the motion model becomes identical to the optical flow model \cite{Horn1981} as used in image registration. 

MRI measurements are acquired in the spectral or spatial frequency domain, called k-space. Therefore, it is convenient to transform the continuity equation (\ref{eq:continuity-equation}) to the spectral domain. Let $\mathcal{F}$ be the $\Ndim$-dimensional spatial Fourier transform operator. It transforms $\fsdit$ into its k-space representation $\hat{\fsdit}(\kpos,t) : \Omega_{kt} \to \mathbb{C}$, such that $\hat{\fsdit} = \mathcal{F}\fsdit$ and $\fsdit = \mathcal{F}^{-1}\hat{\fsdit}$. Here, $\Omega_k \subseteq \mathbb{R}^\Ndim$ is the spectral domain in k-space, and ${\Omega_{kt} = \Omega_k \times \Omega_t}$. Similarly, the spectral velocity field $\hat{\vel}(\kpos,t) : \Omega_{kt} \to \mathbb{C}^\Ndim$ is obtained by applying $\mathcal{F}$ to each component of $\vel$, such that $\hat{\velit}_\idim = \mathcal{F}\velit_\idim$ for $\idim=1,\cdots,\Ndim$. Note that an undersampled and noisy version of $\hat{\fsdit}$ is measured during an MRI scan.

Using the properties of the Fourier transform, (\ref{eq:continuity-equation}) can be converted to the spectral domain: 

\begin{equation}
\label{eq:motion-model-spectral}
    \frac{\partial \hat{\fsdit}}{\partial t} + 2 \pi i \sum_{\idim=1}^\Ndim k_\idim (\hat{\fsdit} * \hat{\velit}_\idim) = 0.
\end{equation}

Here, $i = \sqrt{-1}$ is the imaginary unit, $k_\idim$ is the spectral coordinate in dimension $\idim$, and the symbol $*$ is a convolution over the $\Ndim$ spectral dimensions. 

The equation above is written in terms of the velocity field, while the dynamical model (Section \ref{subsec:dyn-model}) is described in terms of the displacement field. Let $\displ(\pos,t) : \Omega_{rt} \to \mathbb{R}^\Ndim$ be the $\Ndim$-dimensional displacement vector field. We will parameterize the displacement field using $\NDOF$ basis functions $\bfcn_\iDOF(\pos) : \Omega_r \to \mathbb{R}^\Ndim$ and corresponding generalized coordinates $\gencoorit_\iDOF(t) : \Omega_t \to \mathbb{R}$ for $\iDOF = 1,\cdots,\NDOF$:

\begin{equation}
   \displ(\pos,t) = \sum_{\iDOF=1}^\NDOF \bfcn_\iDOF(\pos)\gencoorit_\iDOF(t).
   \label{eq:displ-field}
\end{equation}

\begin{figure}[!t]
    \centerline{\includegraphics[width=\columnwidth]{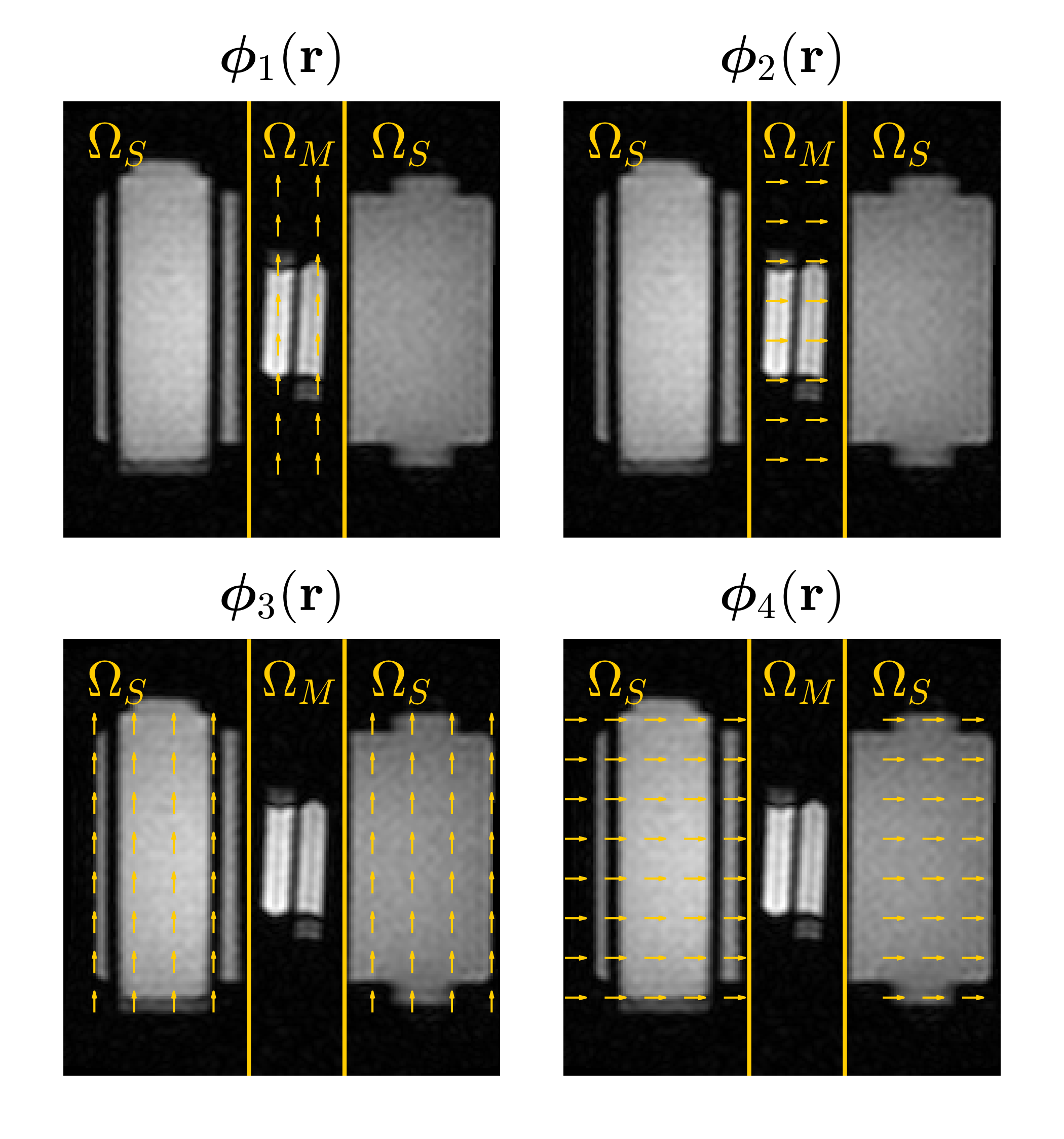}}
    \caption{The four basis functions used during the reconstruction for $\orient = 0\degree{}$. Two basis functions are defined in the moving compartment ($\Omega_M$, top row), the other two in the stationary compartment ($\Omega_S$, bottom row). For each compartment, one basis function points in the vertical direction ($x$, left column), the other in the horizontal direction ($y$, right column). The coefficients of all four basis functions are estimated during the reconstruction, but only those in the moving compartment should be nonzero.}
    \label{fig:basis-functions}
\end{figure}

By using the basis functions $\bfcn_\iDOF$, prior information about the motion field can be introduced, thereby reducing the number of unknown variables. For example, if the motion field is smooth, splines can be used as basis functions. Alternatively, the basis functions can be piecewise constant if the domain $\Omega_r$ can be separated into different compartments. The latter approach will be used in this work (Section \ref{subsec:recon} and Fig. \ref{fig:basis-functions}).

The velocity field and the displacement field are related through the material derivative (${\vel = \frac{D\displ}{Dt} = \frac{\partial\displ}{\partial t} + \vel \cdot \nabla \displ}$). In this work we will limit ourselves to piecewise constant basis functions. Therefore, $\nabla \displit_\idim = \mathbf{0}$ for all $\idim$, and thus $\vel = \frac{\partial \displ}{\partial t}$.
Combining (\ref{eq:motion-model-spectral}) and (\ref{eq:displ-field}) gives:

\begin{equation}
\label{eq:motion-model-spectral-displ}
    \frac{\partial \hat{\fsdit}}{\partial t} + 2 \pi i \sum_{\idim=1}^\Ndim \sum_{\iDOF=1}^\NDOF k_\idim (\hat{\fsdit} * \hat{\bfcnit}_{\iDOF,\idim)} \frac{d \gencoorit_\iDOF}{dt} = 0,
\end{equation}

\noindent where $\hat{\bfcnit}_{\iDOF,\idim} = \mathcal{F}\bfcnit_{\iDOF,\idim}$. 

During an MRI scan, the continuous time-resolved k-space $\hat{\fsdit}$ is sampled at a discrete number of sample points. The k-space domain $\Omega_k$ is discretized to $\Nfull$ points, and $\Omega_t$ is discretized to $\Ntime$ time instances. When all discrete points of $\hat{\fsdit}$ are put into one vector, we get $\fsd \in \mathbb{C}^{\Nfull\Ntime}$. Similarly, we can discretize the generalized coordinates over time to get $\gencoor \in \mathbb{R}^{\NDOF\Ntime}$. Note that $\fsd$ cannot be measured directly at a high spatio-temporal resolution. Instead, a highly undersampled and noisy version of $\fsd$ is acquired (see Section \ref{subsec:data-model}).

After discretization, the motion model term $\MM$ is the squared $L^2$-norm of the residual of (\ref{eq:motion-model-spectral-displ}), which is a function of $\fsd$ and $\gencoor$:

\begin{equation}
    \label{eq:motion-model-discrete}
    \MM(\fsd,\gencoor) = \frac{1}{2} \bigl\| D_t\fsd + 2\pi i \sum_{\idim=1}^\Ndim K_\idim C(\fsd, D_t \bfcnmat_\idim\gencoor)\bigr\|_2^2.
\end{equation}

Here, $D_t$ is the first-order temporal finite difference operator, $K_\idim$ is a diagonal matrix that performs multiplication with the k-space coefficients $k_\idim$, $C$ is a bilinear map that performs the spatial convolution, and $\bfcnmat_\idim$ is a block diagonal matrix whose columns contain the spectral basis functions $\hat{\bfcnit}_{\iDOF,\idim}$.

Note that the term inside the norm in (\ref{eq:motion-model-discrete}) is bilinear in $\fsd$ and $\gencoor$, making $\MM(\fsd, \gencoor)$ a biconvex function.

\subsection{Dynamical Model}
\label{subsec:dyn-model}

If the measured data were fully sampled and free of noise, the coefficients of the motion field could be retrieved directly from the measured data. However, a fully-sampled k-space at a high spatial and temporal resolution is not feasible.

To overcome this issue, the dynamical model is added to constrain the solution of the motion field. This model provides additional information about the expected type of motion. In general, the dynamical model is a PDE constructed from the balance of momentum (Newton's second law) combined with a material's constitutive relation, like Hooke's law for linear elasticity. This constitutive model contains the dynamical parameters, which can either be provided a priori or can be estimated from the data to provide additional information. Examples of dynamical parameters are stiffness and viscosity. 

For now, we will limit ourselves to dynamical models that are only time-dependent, and can thus be written as an ordinary differential equation (ODE). All spatial dependencies are captured by the basis functions $\bfcn_\iDOF$. In this work, we use a linear elastic model with an external activation for each generalized coordinate $\gencoorit_\iDOF$:

\begin{equation}
    \label{eq:dyn-model}
    \frac{d^2\gencoorit_\iDOF}{d t^2} + \damping \frac{d\gencoorit_\iDOF}{dt} + \stiffness \gencoorit_\iDOF = \forceit_\iDOF \qquad \forall \iDOF,
\end{equation}

\noindent with $\damping \in \mathbb{R}_+$ the damping coefficient in Ns/m, $\stiffness \in \mathbb{R}_+$ the elastic stiffness in N/m, and ${\forceit_\iDOF(t) : \Omega_t \to \mathbb{R}}$ the activation force in N for the $\iDOF$-th degree of freedom. 

In our experiments, we set $\damping$ to a particular value, which is fixed during the reconstruction. Also note the absence of the mass in the first term of (\ref{eq:dyn-model}). Since the equation can be scaled by an arbitrary constant, we regard all quantities as mass-normalized. Thus, $\stiffness$ remains as the only unknown dynamical parameter.

Despite its simplicity, this model can be used to describe muscle dynamics in certain tasks \cite{Winters1987MuscleComplexity}, or it can act as a linear approximation to more complex dynamical systems. In this case, $\forceit_\iDOF$ can be interpreted as the internal muscular activation caused by a neuronal input, or as an external loading, depending on the application.

The dynamical model term $\DM$ takes the residual of (\ref{eq:dyn-model}) and takes its squared $L^2$-norm summed over the degrees of freedom. After discretization, this can be written as a function of $\gencoor$, $\stiffness$, and $\force$:

\begin{equation}
    \label{eq:dyn-model-discrete}
    \DM(\gencoor,\stiffness,\force) = \frac{1}{2} \bigl\| \left(D_{tt} + \damping D_t + \stiffness I\right) \gencoor - \force \bigr\|_2^2.
\end{equation}

$D_{t}$ and $D_{tt}$ are the first- and second-order temporal finite difference operators, $I$ is the $\NDOF\Ntime \times \NDOF\Ntime$ identity operator, and $\force \in \mathbb{R}^{\NDOF\Ntime}$ is the discretized activation force.

The term inside the norm of (\ref{eq:dyn-model-discrete}) is linear in $\gencoor$ and $\force$ for a fixed value of $\stiffness$, and it is linear in $\stiffness$ and $\force$ for a fixed value of $\gencoor$. This makes $\DM(\gencoor, \stiffness, \force)$, like $\MM(\fsd, \gencoor)$ in (\ref{eq:motion-model-discrete}), a biconvex function.

\subsection{Data Consistency}
\label{subsec:data-model}

Sampling k-space during an MRI scan is not instantaneous. If all $\Nfull$ samples are to be acquired at each time instance, the time step $\Delta t$ between two time instances becomes too large to capture motion at a sub-second timescale. Therefore, the measured k-space data $\measd \in \mathbb{C}^{\Nmeas\Ntime}$ will be an undersampled version of $\fsd$, with $\Nmeas$ measured samples per time instance, and $\Nmeas$ much smaller than the number of samples in the fully-sampled data ($\Nmeas \ll \Nfull$). In practice, $\Nmeas$ can be the number of samples during a single repetition time (TR), or multiple readouts can be grouped together if the repetition time is small enough. The sampling pattern used during the experiment is given by $\encop \in \{0,1\}^{\Nmeas\Ntime \times \Nfull\Ntime}$, such that $\encop\fsd=\measd$ in the absence of noise. 

In our previous work \cite{VanRiel2022Spectro-DynamicScale}, we used the measurements $\measd$ directly in the motion model. However, this created two problems. First of all, the motion model contains a convolution between the k-space data and the velocity field. This convolution can only be evaluated when the k-space is fully sampled at each time instance. In practice, the k-space of each time instance will be heavily undersampled if a high temporal resolution is required. The data points that are not sampled leave gaps in k-space, while these data points are required to calculate the convolution. In \cite{VanRiel2022Spectro-DynamicScale}, this issue was circumvented by only considering motion fields that vary spatially in the readout direction and are constant in all other directions, resulting in a 1D convolution in the fully-sampled readout direction. However, for more general motion fields or more complex acquisition trajectories, this approach is no longer possible.

Furthermore, $\measd$ will invariably contain measurement noise. By plugging $\measd$ directly into the motion model, this noise gets scaled by the different terms in (\ref{eq:motion-model-discrete}). For example, the high spatial frequency noise is amplified through the operator $K_\idim$. This will result in a biased reconstruction of the motion field.

The first limitation is addressed by including the full time-resolved k-space, including the data points that were not sampled, as an additional variable in the reconstruction. This increases the number of unknown variables, but gives the reconstruction more flexibility. By using a specific interleaved sampling pattern, the information coming from the sampled data at a certain time instance is shared through the temporal derivatives in the motion model to the time instances where those data points are not sampled. That way, the gaps in the time-resolved k-space can be filled in, allowing the convolution to be evaluated for more general motion patterns. 

Instead of using a hard data consistency constraint in the form of $\encop\fsd - \measd = \mathbf{0}$, we will use a relaxation term in the reconstruction that minimizes the error of this equation:

\begin{equation}
    \DC(\fsd) = \frac{1}{2} \|\encop\fsd-\measd\|_2^2.
\end{equation}

This relaxation term allows the reconstructed data $\fsd$ to deviate slightly from the measured data $\measd$. We expect that this deviation can capture the measurement noise and other measurement imperfections, thereby reducing the bias in the motion field caused by these differences.

\begin{figure}[!t]
    \centerline{\includegraphics[width=\columnwidth]{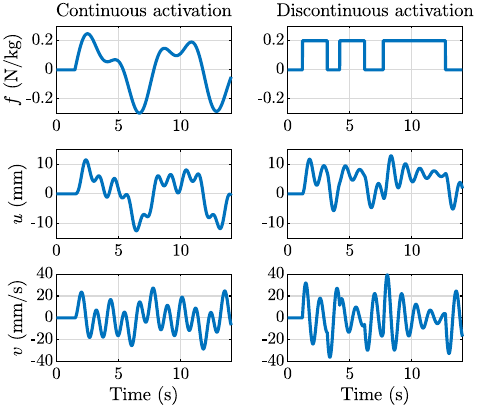}}
    \caption{Simulated inputs for the continuous (left) and discontinuous (right) experiments. On the top row, the activation forces are shown for both cases. The middle and bottom rows show the displacements and velocities respectively, as obtained by numerically solving (\ref{eq:dyn-model}). Only the two small tubes moved along a single direction, while the water compartment and the large cylinder remained stationary.}
    \label{fig:activations}
\end{figure}

\begin{figure}[!t]
    \centerline{\includegraphics[width=\columnwidth]{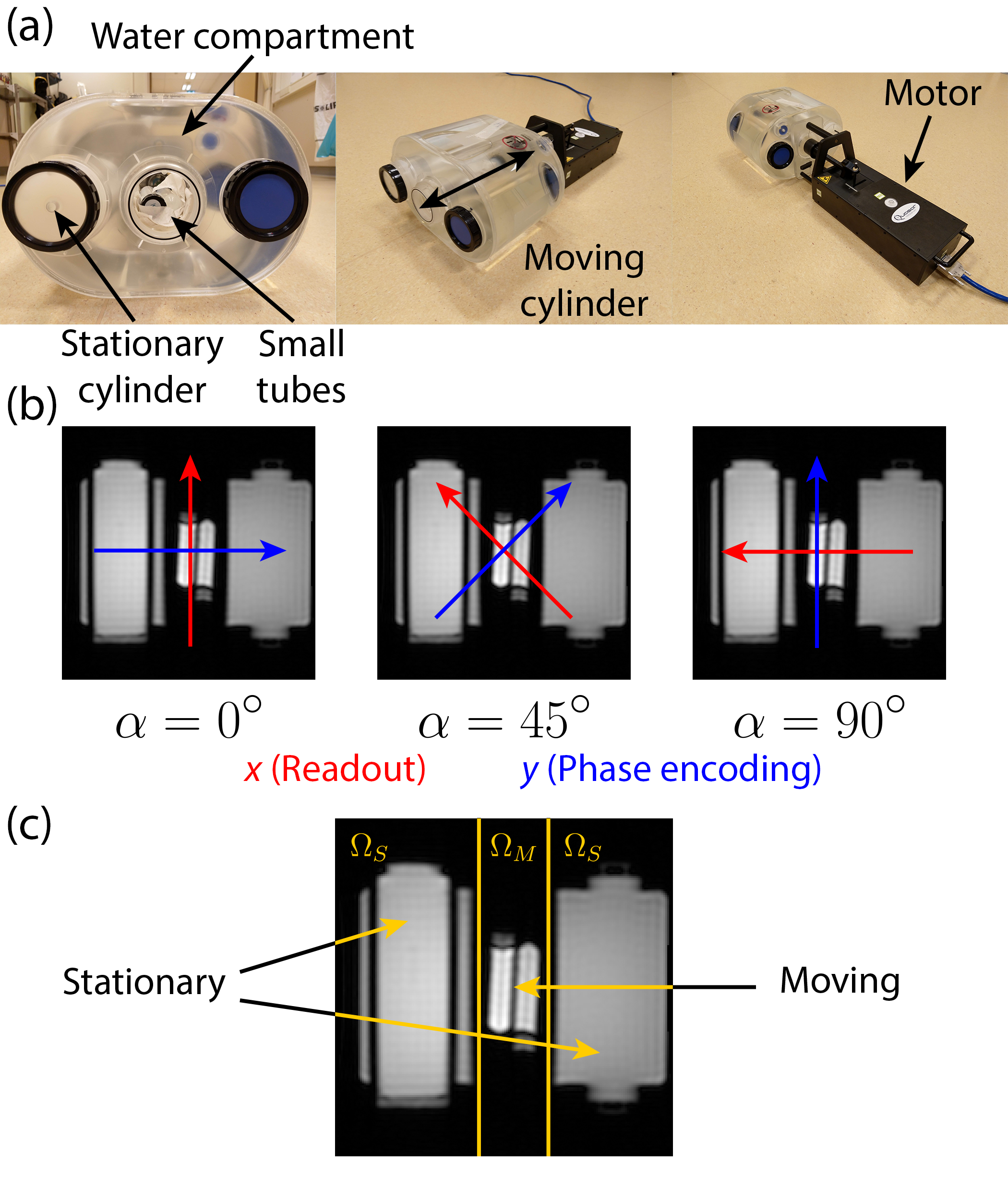}}
    \caption{(a) The Quasar MRI\textsuperscript{4D} Motion Phantom. A large water compartment surrounds a stationary gel-filled tube and a cylinder that is actuated by an MR-compatible piezoelectric motor. Inside the moving cylinder, two smaller gel-filled tubes were placed. (b) Stationary MR images of the phantom setup, acquired with the same scan parameters as those used during the dynamic experiments. The red and blue arrows indicate the readout and phase encode directions respectively for the three different readout orientations ($\orient =$ 0\degree{}, 45\degree{}, and 90\degree{}). These directions correspond to the $x$- and $y$-directions respectively in the reconstructed motion. (c) A stationary MR image, where the two yellow lines indicate the division between the moving compartment ($\Omega_M$) and the stationary compartment ($\Omega_S$).}
    \label{fig:phantom}
\end{figure}

\subsection{Regularization}

By treating $\gencoor$, $\stiffness$ and $\force$ as unknown variables, we have created an ill-posed reconstruction problem. Any error in the dynamical model given by (\ref{eq:dyn-model}) can be compensated by changing its right-hand side, resulting in unfeasible solutions for $\force$. 

Therefore, we added a fourth term $\FR$ to the reconstruction. This term acted as a regularization of the activation force and depends on the prior knowledge that is available about the shape of the activation force. For example, if $\forceit_\iDOF$ is continuous, a smoothness regularization can be used. Alternatively, if $\forceit_\iDOF$ is known to be piecewise constant, a regularization function based on the temporal Total Variation (TV) is more suitable. Both regularization terms can be discretized using the second- and first-order finite difference operators $D_{tt}$ and $D_{t}$ respectively, resulting in:

\begin{equation}
    \label{eq:force-reg-smooth}
    \FR_\text{S}(\force) = \frac{1}{2} \| D_{tt} \force \|_2^2,
\end{equation}

\begin{equation}
    \label{eq:force-reg-TV}
    \FR_\text{TV}(\force) = \| D_t \force \|_1.
\end{equation}

\subsection{Iterative Reconstruction}
\label{subsec:iter-recon}

Combining all four components, the extended Spectro-Dynamic MRI framework aims to solve the following optimization problem:

\begin{equation}
    \label{eq:spect-dyn-minimization}
    \min_{\fsd,\gencoor,\stiffness,\force} \MM(\fsd,\gencoor) + \lDM \DM(\gencoor,\stiffness,\force) + \lDC \DC(\fsd) + \lFR \FR(\force).
\end{equation}

The parameters $\lDM$, $\lDC$, and $\lFR$ determine the trade-off between the residuals of the different models. 

The minimization in (\ref{eq:spect-dyn-minimization}) must be solved for four different parameters: $\fsd$, $\gencoor$, $\stiffness$, and $\force$, for a total of ${\Nfull\Ntime+2\NDOF\Ntime+\Ndynpar}$ unknowns. It is a multiconvex optimization problem, for which several optimization algorithms exist. In this work, we solve (\ref{eq:spect-dyn-minimization}) with an iterative scheme called Block Coordinate Descent (BCD) \cite{Xu2013ACompletion} because it is easy to implement and interpret the results. In every step of BCD, one subset of the unknown variables is fixed, while problem is solved for the remaining variables using convex optimization. This is done most efficiently by fixing a minimal number of variables \cite{Shen2017DisciplinedProgramming}. In the case of (\ref{eq:spect-dyn-minimization}), these minimal fixed sets are $\{\fsd, \stiffness\}$ and $\{\displ\}$. The resulting iterative optimization is given by Algorithm \ref{alg:iterative-optimization}. 

\begin{figure}[!t]
    \centerline{\includegraphics[width=\columnwidth]{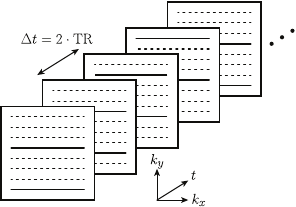}}
    \caption{Sampling pattern as used during the dynamic experiments. The readout direction is $k_x$, while $k_y$ is the phase encode direction. Only eight of the 64 phase encode lines are visualized. Each time instance contains two readout lines, for an effective temporal resolution of two times the repetition time (TR). By distributing the readout lines evenly over the k-space, information about different spatial frequencies is acquired during each time instance.}
    \label{fig:sampling-pattern}
\end{figure}

\begin{algorithm}[H]
\caption{Spectro-Dynamic MRI Reconstruction.}\label{alg:iterative-optimization}
\begin{algorithmic}[1]
\STATE $\fsd^0,\gencoor^0,\stiffness^0,\force^0 \gets 0$
% \STATE $\fsd^0 \gets \operatorname*{arg\,min}_{\fsd} \lMM \frac{1}{2} \|D_t\fsd\|_2^2 + \lDC \DC(\fsd)$
\FOR{$k=1$ to $K$}
\STATE $\displaystyle \fsd^{k} \gets \operatorname*{arg\,min}_{\fsd} \MM(\fsd,\gencoor^{k-1}) + \lDC \DC(\fsd)$
\label{alg:line:min-m}
\label{alg:line:min1}
\STATE $\begin{aligned} \displaystyle \gencoor^{k},\bar{\force}^{k} \gets \operatorname*{arg\,min}_{\gencoor,\force} \Bigl[ & \MM(\fsd^{k},\gencoor) + \lDM \DM(\gencoor,\stiffness^{k-1}, \force) \\ &+ \lFR \FR(\force)\Bigr] \end{aligned}$
\label{alg:line:min-q}
\label{alg:line:min2}
\STATE $\displaystyle \stiffness^{k},\force^{k} \gets \operatorname*{arg\,min}_{\stiffness,\force} \lDM \DM(\gencoor^{k},\stiffness,\force) + \lFR \FR(\force)$
\label{alg:line:min-kappa}
\label{alg:line:min-theta} % To prevent errors when using latexdiff
\label{alg:line:min3}
\ENDFOR
\end{algorithmic}
\end{algorithm}

Note that line \ref{alg:line:min-q} in Algorithm \ref{alg:iterative-optimization} is a convex optimization with $\fsd$ and $\stiffness$ fixed, while lines \ref{alg:line:min-m} and \ref{alg:line:min-kappa} together form a convex optimization with $\gencoor$ fixed. This last minimization problem over $\fsd$, $\stiffness$ and $\force$ is separable into two independent minimization problems: one over $\fsd$ (line \ref{alg:line:min-m}), and one over $\stiffness$ and $\force$ (line \ref{alg:line:min-kappa}). The ordering of the subproblems is chosen such that the influence of the initial values is minimal. The activation force $\force$ is optimized twice in each iteration, therefore $\bar{\force}^k$ is a nuisance variable only used to improve the solution of $\displ$.

When the smooth regularization function (\ref{eq:force-reg-smooth}) is used, all three subproblems (lines \ref{alg:line:min1}-\ref{alg:line:min3} in Algorithm \ref{alg:iterative-optimization}) can be solved using linear least squares solvers. In the case of the Total Variation regularizer (\ref{eq:force-reg-TV}), lines \ref{alg:line:min-q} and \ref{alg:line:min-kappa} are still convex, but no longer smooth. In this case, these are solved using the Alternating Direction Method of Multipliers (ADMM) \cite[pp.~43--44]{Boyd2011DistributedMultipliers}.

%% file: 3_Methods.tex
\subsection{Activation Functions}

We used two different kinds of activations to observe different kinds of dynamics in our system. First, a smooth continuous activation function was constructed by adding two sinusoidal functions with frequencies of 0.15 Hz and 0.33 Hz (see Fig. \ref{fig:activations}, top left). For this activation, the reconstruction was regularized by $\FR_\text{S}$. The next experiment used a discontinuous activation, where the activation switched between an 'off' and an 'on' state (Fig. \ref{fig:activations}, top right). For this experiment, $\FR_\text{TV}$ was used as a regularization function. 

The differential equation of motion (\ref{eq:dyn-model}) was solved using $\stiffness = \textrm{30 N/m}$ and the initial condition ${\gencoorit(0) = 0}$, where $\forceit(t)$ is given by one of the activations in Fig. \ref{fig:activations}. We used a damping coefficient of $\damping = \textrm{0 Ns/m}$ for the continuous activation, and $\damping = \textrm{1 Ns/m}$ for the discontinuous activation. This value was fixed during the reconstruction.

\subsection{Phantom Setup}

We used the Quasar MRI\textsuperscript{4D} Motion Phantom (Modus Medical Devices Inc., London, ON, Canada) to acquire dynamic experimental data. This phantom (Fig. \ref{fig:phantom}(a)) contains an MR-compatible piezoelectric motor, which drives a cylindrical moving compartment inside a stationary compartment filled with water. Inside the moving cylinder, two gel-filled tubes (TO5, Eurospin II test system, Scotland) were placed. A second, static cylinder filled with gel was inserted next to the moving cylinder. The motion fields were piecewise constant, which means that the motion was locally rigid. As such, they can be described by four degrees of freedom (two for the moving compartment and two for the stationary compartment). The simulated motion profile (Fig. \ref{fig:activations}, middle row) was used as position setpoint in the phantom's motion control software. 

\subsection{Data Acquisition}

\begin{figure*}[!t]
    \centerline{\includegraphics[width=\textwidth]{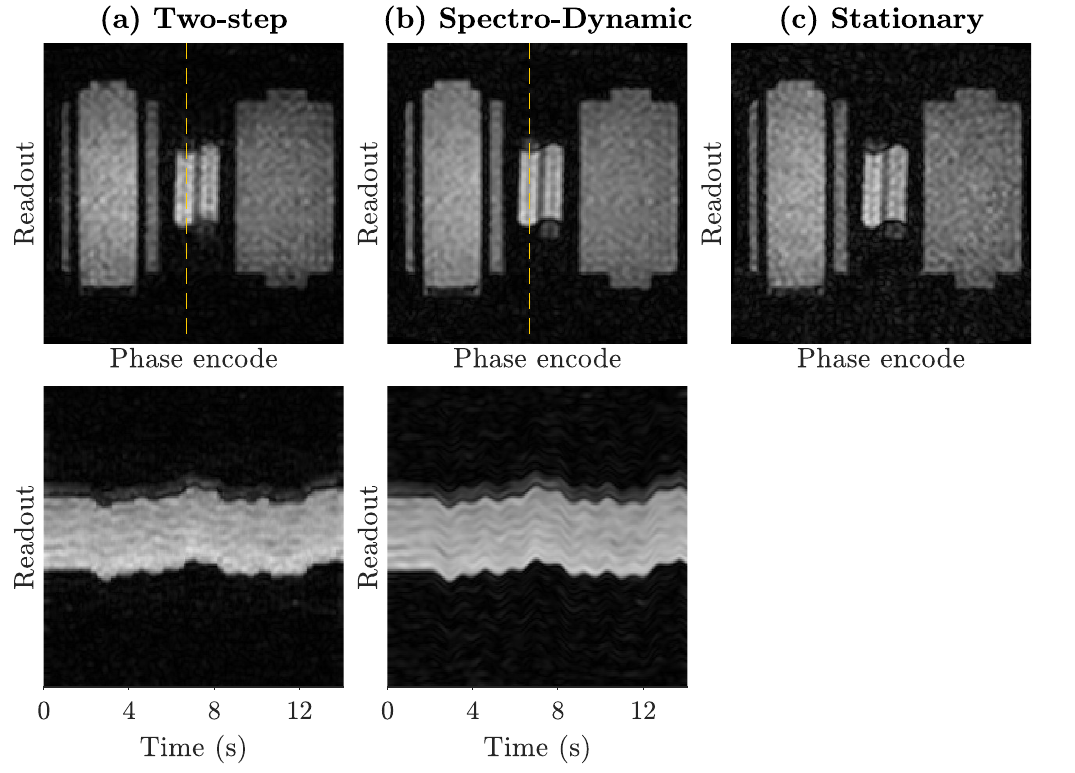}}
    \caption{Images reconstructed using (a) the two-step reconstruction, (b) the result of the Spectro-Dynamic MRI reconstruction, and (c) stationary data. The readout direction was parallel to the direction of motion ($\orient = 0\degree{}$). The bottom row shows one line in the image (indicated by the dashed yellow line) over time. Note the removal of the motion blurring in the Spectro-Dynamic MRI reconstruction. A video showing all frames of the time-resolved Spectro-Dynamic MRI reconstruction is available online as Supplementary Video S\ref{svid:recon-cont-0}.}
    \label{fig:result-images}
\end{figure*}

The motion phantom was placed in a 1.5T MRI scanner (Ingenia, Philips Healthcare, Best, The Netherlands). Both stationary and dynamic data were acquired with a spoiled gradient echo sequence (TR $=$ 5.5 ms, TE $=$ 2.2 ms, Flip angle $=$ 9\degree{}). The field of view was a 320 mm $\times$ 320 mm coronal slice with a thickness of 5 mm. The phantom was leveled to prevent through-slice motion. To achieve a homogeneous receive sensitivity, the body coil was used for data acquisition. Data sampling was performed on a 64 $\times$ 64 matrix using an interleaved pattern (Fig. \ref{fig:sampling-pattern}). Thus, the acquired information at each time instance was distributed over k-space. The sampling pattern was repeated 40 times, for a total of 2560 readouts, and a total acquisition time of 14 seconds.

To show that the extended Spectro-Dynamic MRI framework can estimate motion along an arbitrary direction, the experiment was repeated three times for both activations. Each time, the readout direction ($k_x$) was rotated respectively by 0\degree{}, 45\degree{}, and 90\degree{} with respect to the direction of motion (Fig. \ref{fig:phantom}(b)).

\subsection{Reconstruction}
\label{subsec:recon}

Four degrees of freedom ($P = \textrm{4}$) were used during reconstruction, for two compartments (the moving compartment $\Omega_M$ and the stationary compartment $\Omega_S$, Fig. \ref{fig:phantom}(c)) and two directions ($x$ and $y$) per compartment. Thus, four piecewise constant basis functions were defined to parameterize the displacement vector field according to (\ref{eq:displ-field}), corresponding with four generalized coordinates (Fig. \ref{fig:basis-functions}):

\begin{equation}
    \begin{aligned}
    \bfcn_1(\pos) &= \begin{cases} (1,0)^T \qquad \forall \pos \in \Omega_M \\ (0,0)^T \qquad \forall \pos \in \Omega_S \end{cases} \\
    \bfcn_2(\pos) &= \begin{cases} (0,1)^T \qquad \forall \pos \in \Omega_M \\ (0,0)^T \qquad \forall \pos \in \Omega_S \end{cases} \\
    \bfcn_3(\pos) &= \begin{cases} (0,0)^T \qquad \forall \pos \in \Omega_M \\ (1,0)^T \qquad \forall \pos \in \Omega_S \end{cases} \\
    \bfcn_4(\pos) &= \begin{cases} (0,0)^T \qquad \forall \pos \in \Omega_M \\ (0,1)^T \qquad \forall \pos \in \Omega_S \end{cases}
    \end{aligned}
    \label{eq:basis-functions}
\end{equation}

Since the moving compartment moves along a straight line, the reconstructed displacements of only one degree of freedom should be nonzero for $\orient = $ 0\degree{} and 90\degree{}. For $\orient = $ 45\degree{}, the displacements of two degrees of freedom should be nonzero and identical.

Multiple successive readouts were grouped together during reconstruction. This was done to increase the amount of available data per time instance while guaranteeing a high temporal resolution. Including more lines in a single time instance reduces the undersampling of k-space, at the cost of a lower temporal resolution. We grouped together every two successive readouts (Fig. \ref{fig:sampling-pattern}). This resulted in an effective temporal resolution of $\Delta t = \textrm{2} \cdot \textrm{TR} = \textrm{11 ms}$ and an undersampling factor of 32 for each time instance. This temporal resolution enables future applications in real-time dynamic cardiac or speech imaging, which require similar frame rates \cite{Nayak2022Real-TimeImaging}.

The reconstruction, as described by Algorithm \ref{alg:iterative-optimization}, was implemented in Matlab (The MathWorks Inc., Natick, MA, USA). The large-scale and sparse subproblem in line \ref{alg:line:min-m} of Algorithm \ref{alg:iterative-optimization} was solved using the iterative linear least-squares solver \texttt{lsqr}. The other two subproblems were solved using Matlab's \texttt{mldivide} (\textbackslash) or an ADMM implementation, depending on $\FR(\force)$. The regularization parameters (${\lDM = \textrm{5.0} \cdot \textrm{10\textsuperscript{6}}}$, ${\lDC = \textrm{1.0} \cdot \textrm{10\textsuperscript{3}}}$, and ${\lFR = \textrm{1.0} \cdot \textrm{10\textsuperscript{3} / 2.0} \cdot \textrm{10\textsuperscript{4}}}$ for $\FR_\text{S}\textrm{ / }\FR_\text{TV}$) were determined empirically and fixed for all reconstructions. These values resulted in comparable contributions of all four model terms to the objective function. After 15 iterations, the updates to the estimated values were negligible (less than 1\%), and the algorithm was terminated. Each reconstruction took approximately 2 hours on a workstation with a 2.90 GHz Intel Xeon W-2102 CPU and 32GB RAM.

The estimated fully-sampled k-space $\fsd$ was transformed into the image domain with a Fast Fourier Transform (FFT) followed by a geometry correction step to correct for gradient nonlinearities. The velocity was calculated using central finite differences from the estimated displacements. Together, the activation force and the velocity are important for understanding muscle physiology \cite{Alcazar2019OnDouble-Hyperbolic}. For the estimated displacements, velocities, and activation forces, the root-mean-square error (RMSE) compared to the ground truth values (Fig. \ref{fig:activations}) was calculated as a measure for the reconstruction accuracy.

\subsection{Two-step Reconstruction}
\label{subsec:two-step-recon}

We compared the proposed Spectro-Dynamic MRI reconstruction to a two-step approach. In the first step, the time-resolved k-space $\fsd$ was reconstructed by only using a temporal smoothness regularization and no information about the motion fields:

\begin{equation}
    \label{eq:temp-smooth-minimization}
    \min_{\fsd} \DC(\fsd) + \lambda \|D_t \fourierop^H \fsd\|_1,
\end{equation}

\noindent with $\fourierop^H$ the inverse Fourier transform operator and $\lambda = \textrm{0.05}$. The minimization in (\ref{eq:temp-smooth-minimization}) was solved using the nonlinear conjugate gradient algorithm \cite{Lustig2007SparseImaging}. Next, the remaining variables $\gencoor$, $\stiffness$, and $\force$ were estimated by iterating over lines \ref{alg:line:min-q} and \ref{alg:line:min-kappa} of Algorithm \ref{alg:iterative-optimization} while keeping $\fsd$ fixed. This approach is referred to as the "two-step" reconstruction.

%% file: 4_Results.tex
\begin{table*}[!t]
    \centering
    \caption{Reconstruction errors and estimated dynamical parameter.}
    \begin{tabular}{cccccccccccc}
        \toprule
        \multicolumn{2}{c}{\textbf{Experiment}} && \multicolumn{4}{c}{\textbf{Two-step}} && \multicolumn{4}{c}{\textbf{Spectro-Dynamic}} \\
        \cmidrule{1-2} \cmidrule{4-7} \cmidrule{9-12}
        \textbf{Activation} & $\orient$ && \textbf{RMSE u} & \textbf{RMSE v} & \textbf{RMSE f} & $\stiffness$* && \textbf{RMSE u} & \textbf{RMSE v} & \textbf{RMSE f} & $\stiffness$* \\
        & (\degree{}) && (mm) & (mm/s) & (N) & (N/m) && (mm) & (mm/s) & (N) & (N/m) \\
        \cmidrule{1-2} \cmidrule{4-7} \cmidrule{9-12}
        Continuous & 0 && 1.38 & 3.55 & 27$\cdot$10\textsuperscript{-3} & 53.8 && 0.24 & 1.00 & 9.5$\cdot$10\textsuperscript{-3} & 31.8 \\
        Continuous & 45 && 1.48 & 3.62 & 14$\cdot$10\textsuperscript{-3} & 58.4 && 0.11 & 0.87 & 7.0$\cdot$10\textsuperscript{-3} & 30.1 \\
        Continuous & 90 && 1.54 & 3.72 & 20$\cdot$10\textsuperscript{-3} & 67.0 && 0.20 & 0.96 & 8.6$\cdot$10\textsuperscript{-3} & 30.4 \\
        Discontinuous & 0 && 1.36 & 4.95 & 38$\cdot$10\textsuperscript{-3} & 28.4 && 0.23 & 1.42 & 17$\cdot$10\textsuperscript{-3} & 28.0 \\
        Discontinuous & 45 && 1.45 & 5.11 & 54$\cdot$10\textsuperscript{-3} & 18.3 && 0.11 & 0.85 & 15$\cdot$10\textsuperscript{-3} & 26.8 \\
        Discontinuous & 90 && 1.38 & 5.07 & 47$\cdot$10\textsuperscript{-3} & 22.3 && 0.19 & 1.36 & 16$\cdot$10\textsuperscript{-3} & 27.4 \\
        \bottomrule \addlinespace[1mm]
        \multicolumn{6}{l}{* Ground truth value: 30.0 N/m}
    \end{tabular}
    \label{tab:results}
\end{table*}

In the dynamic image series estimated in the two-step reconstruction, residual motion blurring was visible (Fig. \ref{fig:result-images}(a)). The images calculated from the reconstructed spatio-temporal k-space data $\fsd$ did not show this blurring (Fig. \ref{fig:result-images}(b)). For validation purposes, we also reconstructed an image from fully-sampled k-space data of a stationary phantom (Fig. \ref{fig:result-images}(c)). Note that Fig. \ref{fig:result-images}(b) and (c) are very similar, although the phantom moved while acquiring data for the Spectro-Dynamic MRI reconstruction (Fig. \ref{fig:result-images}(b)). The reconstructed image series had a temporal resolution of 11 ms (Supplementary Video S\ref{svid:recon-cont-0}).

The estimated displacements, velocities, and activation forces were close to their respective ground truth values (Fig.~\ref{fig:result-dyn-cont} and Table \ref{tab:results}). Additionally, the estimated dynamical parameter ($\stiffness$) was very close to its ground truth value of 30 N/m (Table \ref{tab:results}).

When rotating the readout direction with respect to the direction of motion, the reconstruction accuracy remained high (Table \ref{tab:results} and Figs. S\ref{sfig:dyn-cont-45} and S\ref{sfig:dyn-cont-90} in the Supplementary Material). In addition, when changing the activation force to a discontinuous 'on/off' function (Fig. \ref{fig:activations}, right column), the reconstructed motion remained accurate (Table \ref{tab:results} and Fig. S\ref{sfig:dyn-discont-0} in the Supplementary Material), although the accuracy of the estimated activation force and dynamical parameter was lower than for the reconstructions with the continuous activation. Again, the accuracy remained similar when the readout direction was rotated (Table \ref{tab:results} and Figs. S\ref{sfig:dyn-discont-45} and S\ref{sfig:dyn-discont-90} in the Supplementary Material).

%% file: 5_Discussion.tex
\begin{figure*}[!t]
    \centerline{\includegraphics[width=\textwidth]{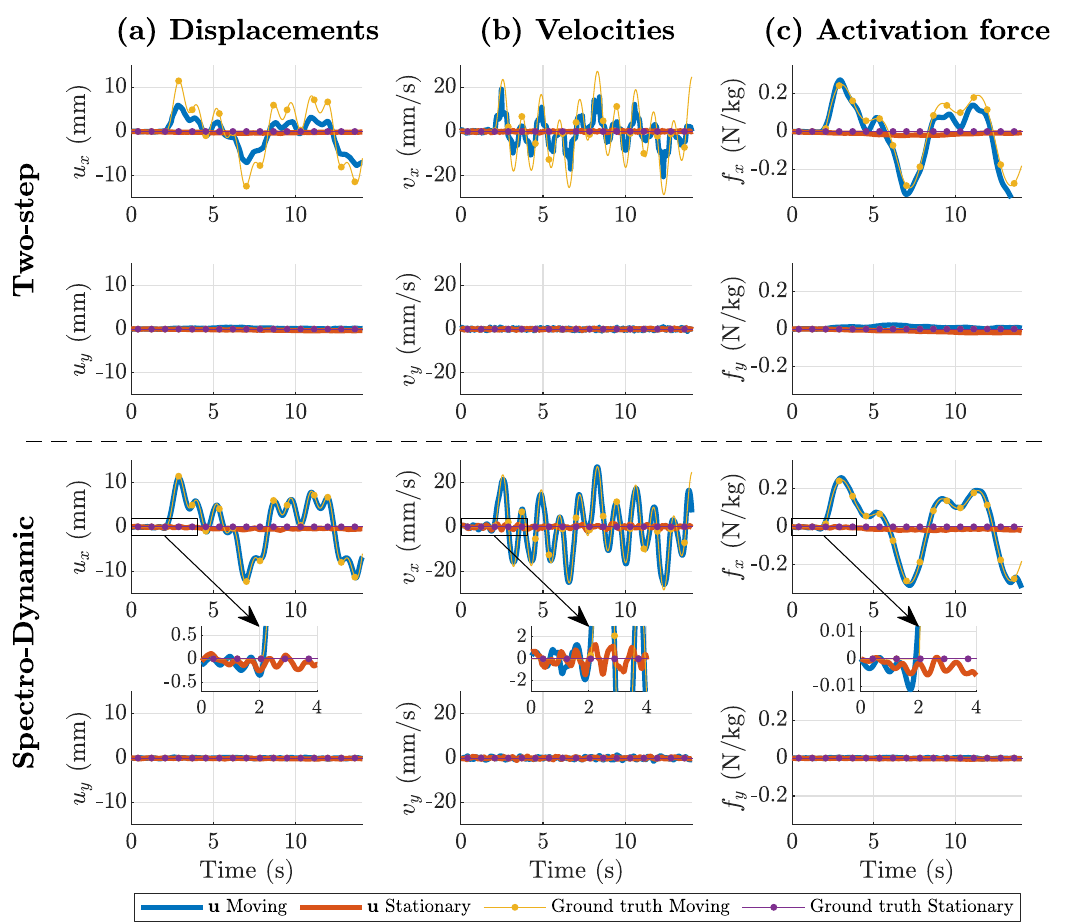}}
    \caption{Estimated (a) displacements, (b) velocities, and (c) activation forces in both directions resulting from the two-step reconstruction (top) and Spectro-Dynamic reconstruction (bottom) for the experiment with a continuous activation function and $\orient = 0\degree{}$. The velocities were calculated from the displacements using finite differences. The motion of both the moving and the stationary compartment was estimated, but only the moving compartment moved along one direction ($x$). The small insets show zoomed versions of the graphs around the point where the motion starts.}
    \label{fig:result-dyn-cont}
\end{figure*}

In this work, we presented an extended reconstruction framework for Spectro-Dynamic MRI that estimates time-resolved images, time-resolved motion fields, dynamical parameters, and the activation force. We formulated an optimization problem in terms of four components: a motion model that relates the displacements to the fully-sampled k-space data, a dynamical model that describes the expected type of motion, a data consistency term that describes the sampling and measurement process, and a regularization term on the activation force. We solved this optimization problem by using an iterative reconstruction algorithm with convex subproblems. Phantom experiments with different kinds of activations showed that the dynamics can be reconstructed accurately at a temporal resolution of 11 ms. The reconstruction did not assume any periodicity in the motion, and the accuracy was independent of the orientation of the motion with respect to sampling pattern.

The strength of our method is the joint reconstruction of the missing k-space data and the motion fields. We have shown the increased accuracy of our proposed reconstruction method compared to a two-step approach, where time-resolved images are reconstructed in the first step, after which the motion is estimated from this data. Since no motion information is available during the first step, a temporal smoothness regularization is used to resolve the high undersampling, which results in motion blurring in the images. Estimating the motion from this blurred data results in a large underestimation of the displacements. In contrast, the iterative Spectro-Dynamic MRI reconstruction leverages the motion information during the estimation of the time-resolved data and vice versa. Thus, the measured information can be shared more efficiently across time, and accurate reconstructions with high acceleration factors per time instance (32 in this work) can be achieved.

The ordering of the subproblems in Algorithm \ref{alg:iterative-optimization} is important. In the first iteration, the optimization for $\fsd$ (with $\gencoor^0 = \mathbf{0}$) results in a smooth temporal interpolation of the missing k-space data. This rough estimate for $\fsd$ is subsequently used to generate a first estimate of the motion fields and activation force. Finally, the dynamical parameters are estimated, while the activation force is refined. Thus, the result after one iteration already resembles the eventual solution. Subsequent iterations refine the solution, bringing it closer to the ground truth values. Choosing a different ordering can result in convergence problems, especially in the first iteration, and increases the sensitivity of the solution to the chosen initial values.

The activation force $\force$ is optimized twice during each iteration of Algorithm \ref{alg:iterative-optimization}. Omitting the estimation of $\force$ in either line \ref{alg:line:min-q} or \ref{alg:line:min-kappa} results in a much slower convergence of $\gencoor$ or $\stiffness$, respectively. Even though the temporary variable $\bar{\force}^k$ is not used explicitly, it gives more freedom during the convex optimization of line \ref{alg:line:min-q}, resulting in a better estimate for $\gencoor$. 

The damping coefficient $\damping$ was fixed during the optimization, since estimating both $\damping$ and $\force$ would lead to complexities in the reconstruction. For more complex and more realistic in vivo applications, some mechanical parameters could be fixed to literature values. Alternatively, the time-dependent activation function could be measured using e.g. ECG or EMG, depending on the application.

Due to the limitations of the motion phantom we used, only one of the four degrees of freedom of the motion was actually nonzero. Since the motion estimation remained accurate when the readout direction was rotated, we expect that motion in any of the four degrees of freedom can be reconstructed accurately.

By using the data consistency term $\DC$ as a penalty term instead of an equality constraint, the noise in the measurements can be captured by the residual of this term. The idea is that this relaxation strategy reduces the noise in $\fsd$ which is used in the motion model $\MM$. This in turn results in a reduced bias of the estimated displacement field. This bias was visible for example in the estimated displacements in the phase encode direction in our previous work \cite{VanRiel2022Spectro-DynamicScale}, which were less accurate than the displacements in the readout direction. In our current results, no such difference was present between the different directions. The addition of the data consistency term in the reconstruction also removed the need for the low-pass filter which was previously applied to the k-space data.

The motion model leverages the reconstructed displacement field to fill the missing k-space data. Other techniques also aim to reconstruct images from undersampled k-space data. Parallel imaging \cite{Pruessmann1999SENSE:MRI, Griswold2002GeneralizedGRAPPA} uses information about the coil sensitivities to estimate the missing data, thereby removing the undersampling artifacts. Compressed sensing \cite{Lustig2007SparseImaging} combines an incoherent undersampling pattern with a sparse regularization to reconstruct an image. In addition, multiple methods have been developed to accelerate dynamic MRI data using temporal correlations in the data, without using the motion fields explicitly \cite{Tsao2012MRITechniques}. These techniques could complement the motion model in Spectro-Dynamic MRI, although this is not done in this work.

Our method requires several extensions before it can be used for in vivo applications. Firstly, through-slice motion violates the assumption of signal conservation used in the derivation of the motion model. Since pure in-plane in vivo motion is rare, we will next investigate a 3D implementation for Spectro-Dynamic MRI. Acquiring 3D data at a high temporal resolution will require undersampling in both phase encode directions, resulting in an even higher undersampling factor than in the current 2D implementation. We expect that carefully choosing the sampling pattern will be important to keep a good trade-off between spatial and temporal resolution. More than two readouts may have to be included in a single time instance. Again, techniques such as parallel imaging or compressed sensing may be used to deal with the increased undersampling in 3D. Random undersampling patterns or non-Cartesian trajectories might also increase the robustness of the reconstructions, especially when using compressed sensing reconstruction approaches. Prior information about the estimated images could also be added to the reconstruction in the form of a smoothness or total variation regularizer.

Secondly, a partial differential equation (PDE) has to replace the ODE used as dynamical model in this work. This PDE allows for the reconstruction of deformable motion fields. For example, the Navier-Cauchy equation can be used to model the behavior of isotropic linear elastic materials. In this case, the non-rigid motion fields become a function of space as well as time. The mechanical parameters might also become spatially dependent when different tissue types or inhomogeneous tissues are considered. We will investigate deformable motion estimation and the use of PDEs as dynamical model in the Spectro-Dynamic MRI framework in the future.

A non-rigid motion field also calls for different basis functions $\bfcn(\pos)$. B-splines \cite{Rohlfing2003Volume-PreservingConstraint} could be used, since motion fields are smooth almost everywhere. Furthermore, a low-rank description of the motion field has been shown to be very effective \cite{Huttinga2020Non-rigidMR-MOTUS}. Choosing adequate basis functions to describe the motion field will be an important step in extending Spectro-Dynamic MRI towards 3D in vivo applications.

Finally, determining constitutive models to describe the dynamic behavior of tissues in vivo is not trivial, and many different (nonlinear) models can be found in the literature \cite{Chagnon2015HyperelasticReview}. Alternatively, data-driven discovery of dynamical systems can be used to learn constitutive relations from the dynamical data itself \cite{Brunton2016DiscoveringSystems, Flaschel2021UnsupervisedLaws}. Discovering dynamical models using Spectro-Dynamic MRI is currently being investigated by the authors \cite{Heesterbeek2023Data-drivenSystem}.

%% file: 6_Conclusion.tex
The extended Spectro-Dynamic MRI reconstruction framework presented in this work allows for time-resolved dynamic MRI reconstructions at a high temporal resolution. In addition, it gives insights into the underlying dynamics of the motion through the dynamical model. The extended framework enables the reconstruction of more general motion fields, regardless of the sampling pattern adopted during the MRI scan.

%% file: 7_Biography.tex
\begin{IEEEbiographynophoto}{Max H. C. van Riel}
received the B.Sc. degree in Biomedical Engineering from Eindhoven University of Technology, Eindhoven, The Netherlands, in 2018, and the M.Sc. degree in Medical Imaging from Eindhoven University of Technology, Eindhoven, The Netherlands in 2020.

In 2019, he was a Student Intern at the New York University School of Medicine for six months. He is currently pursuing a Ph.D. degree at the UMC Utrecht, Utrecht, The Netherlands. In his research, he focuses on characterizing dynamical systems at a high temporal resolution using MRI.
\end{IEEEbiographynophoto}

\begin{IEEEbiographynophoto}{Tristan van Leeuwen}
received the M.Sc. degree in Computational Science from Utrecht University, Utrecht, The Netherlands, in 2006, and the Ph.D. degree in Geophysics from Delft University of Technology, Delft, The Netherlands, in 2010.

He was a Postdoctoral Fellow at the University of British Columbia from 2010-2013, and at the Centrum voor Wiskunde en Informatica (CWI) from 2013-2014. He was an Assistant Professor at the Mathematical Institute of Utrecht University, Utrecht, The Netherlands. Currently, he leads the Computational Imaging Group at CWI in Amsterdam, The Netherlands. His research interests include: inverse problems, computational imaging, tomography and numerical optimisation.
\end{IEEEbiographynophoto}

\begin{IEEEbiographynophoto}{Cornelis A. T. van den Berg}
received the M.Sc. degree in Applied Physics from University of Twente, Enschede, The Netherlands, in 2000, and the Ph.D. degree from Utrecht University, Utrecht, The Netherlands, in 2006.

At present, he is a Full Professor at the UMC Utrecht, Utrecht, The Netherlands, where he leads the Computational Imaging Group. Here, he is (co)leading various research projects on MRI methods \& hardware for MRI guided radiotherapy, new MRI diagnostics using MR-STAT technology, automated imaging workflows for radiotherapy using deep learning and his first love on electromagnetics of MRI.
\end{IEEEbiographynophoto}

\begin{IEEEbiographynophoto}{Alessandro Sbrizzi}
received the B.Sc. and M.Sc. degrees in Mathematics in 2009 and the Ph.D. degree in 2013, all from Utrecht University, Utrecht, The Netherlands. 

He is currently an Associate Professor at the Computational Imaging group of the UMC Utrecht, Utrecht, The Netherlands. His research focuses on fast multi-parametric MRI (in particular MR-STAT), real-time motion-estimation (MR-MOTUS technique), radiofrequency pulse design and the application of machine learning in MRI. 
\end{IEEEbiographynophoto}

%% file: supplementary.tex
\renewcommand{\thesection}{S.\Roman{section}} 
\renewcommand{\thesubsection}{\thesection\Alph{subsection}}

\makeatletter
\makeatletter \renewcommand{\fnum@table}
{\tablename~S\thetable}
\makeatother

\makeatletter
\makeatletter \renewcommand{\fnum@figure} {\figurename~S\thefigure} 
\makeatother

\renewcommand{\theequation}{S\arabic{equation}}

\setcounter{figure}{0}

\begin{center}
\Huge Time-Resolved Reconstruction of Motion, Force, and Stiffness using Spectro-Dynamic MRI

Supplementary material

\vspace{10pt}

\normalsize Max H. C. van Riel\textsuperscript{1}, Tristan van Leeuwen\textsuperscript{2}, Cornelis A. T. van den Berg\textsuperscript{1}, and Alessandro Sbrizzi\textsuperscript{1}

\vspace{5pt}

\footnotesize \textsuperscript{1}Department of Radiotherapy, Computational Imaging Group for MR Diagnostics and Therapy, UMC Utrecht, Utrecht, The Netherlands

\textsuperscript{2}Mathematical Institute, Utrecht University, Utrecht, The Netherlands
\end{center}

% \author{ \\
% \footnotesize{}

\begin{figure*}[!h]
    \centerline{\includegraphics[width=\textwidth]{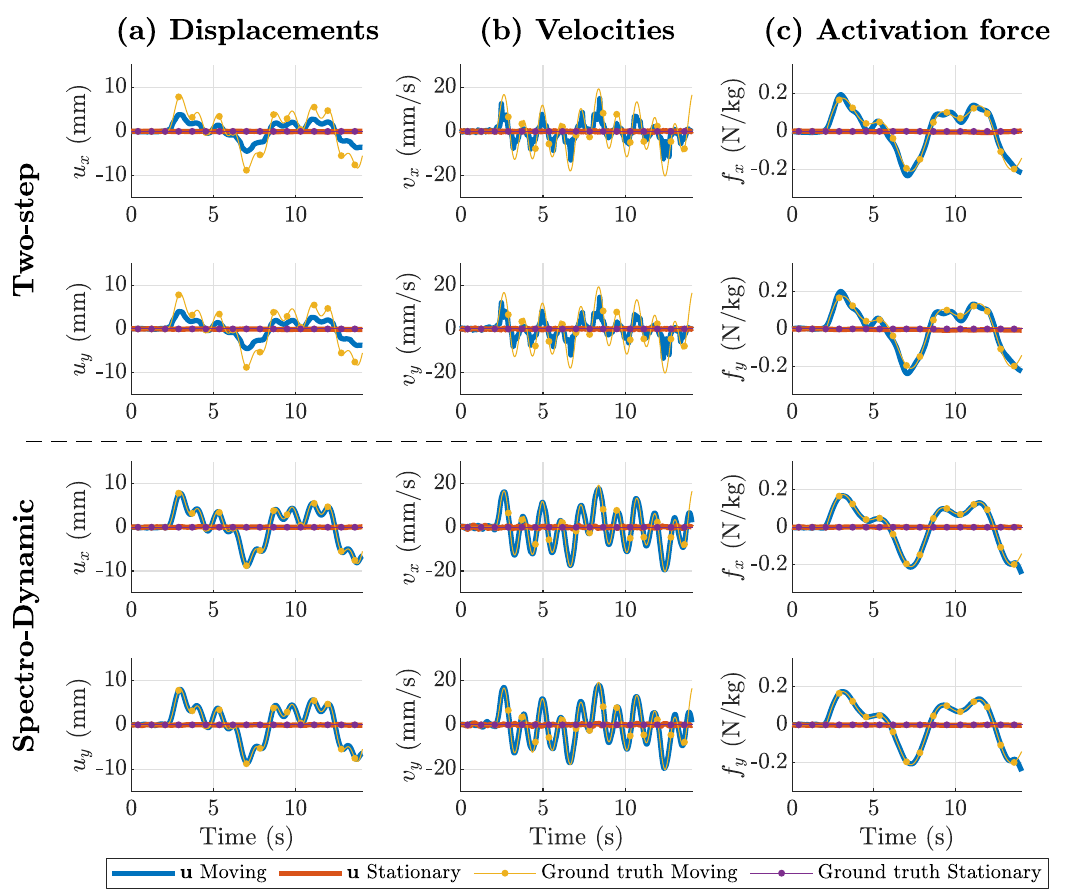}}
    \caption{Estimated (a) displacements, (b) velocities, and (c) activation forces in both directions resulting from the two-step reconstruction (top) and Spectro-Dynamic reconstruction (bottom) for the experiment with a continuous activation function and $\orient = 45\degree{}$. The velocities were calculated from the displacements using finite differences. The motion of both the moving and the stationary compartment was estimated, but only the moving compartment moved along one direction. Since $\orient=45\degree{}$, this motion was separated over the $x$- and $y$-directions.}
    \label{sfig:dyn-cont-45}
    \label{sfig:dyn-first}
\end{figure*}

\begin{figure*}[!t]
    \centerline{\includegraphics[width=\textwidth]{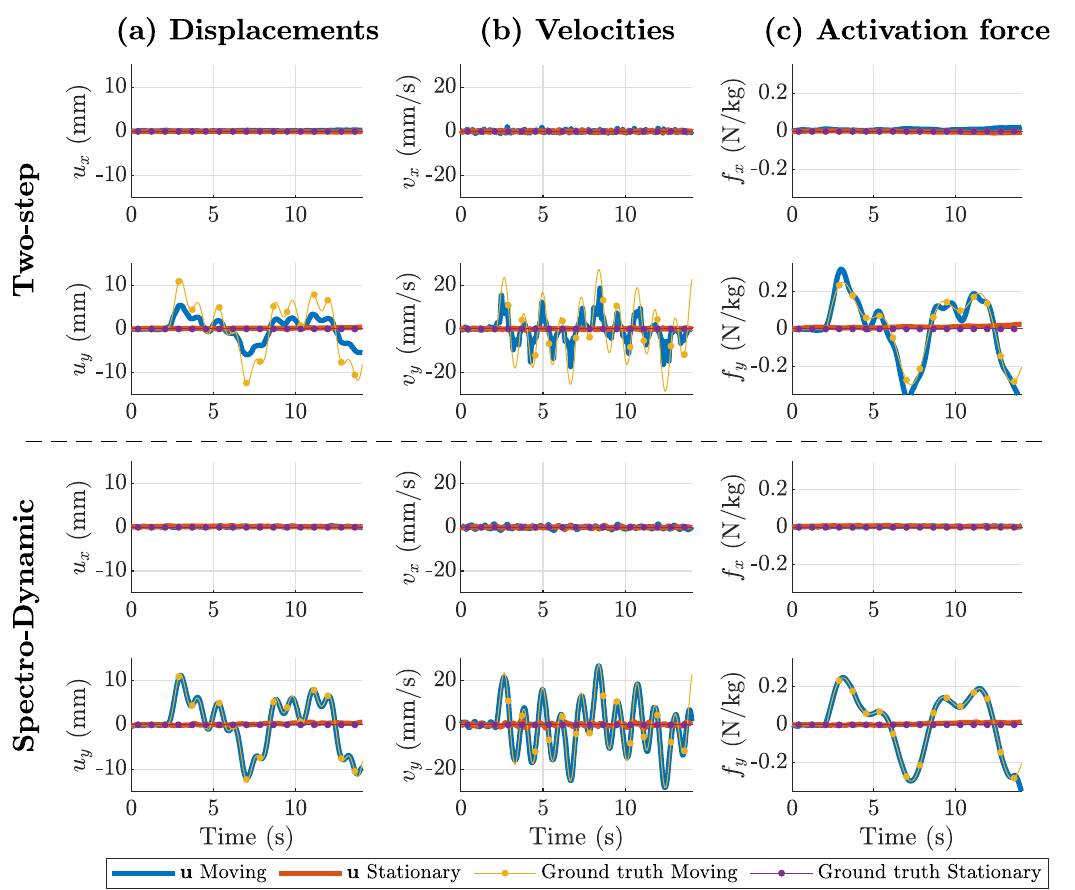}}
    \caption{Estimated (a) displacements, (b) velocities, and (c) activation forces in both directions resulting from the two-step reconstruction (top) and Spectro-Dynamic reconstruction (bottom) for the experiment with a continuous activation function and $\orient = 90\degree{}$. The velocities were calculated from the displacements using finite differences. The motion of both the moving and the stationary compartment was estimated, but only the moving compartment moved along one direction ($y$).}
    \label{sfig:dyn-cont-90}
\end{figure*}

\begin{figure*}[!t]
    \centerline{\includegraphics[width=\textwidth]{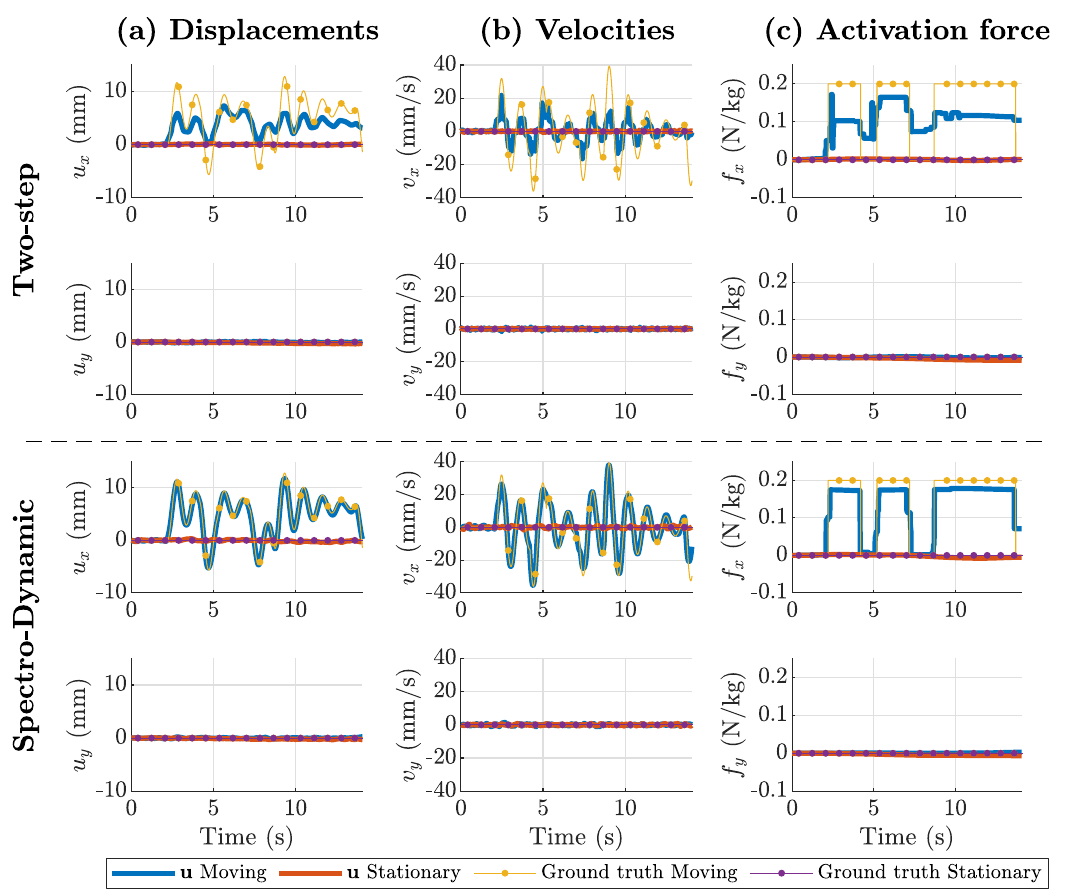}}
    \caption{Estimated (a) displacements, (b) velocities, and (c) activation forces in both directions resulting from the two-step reconstruction (top) and Spectro-Dynamic reconstruction (bottom) for the experiment with a discontinuous activation function and $\orient = 0\degree{}$. The velocities were calculated from the displacements using finite differences. The motion of both the moving and the stationary compartment was estimated, but only the moving compartment moved along one direction ($x$).}
    \label{sfig:dyn-discont-0}
\end{figure*}

\begin{figure*}[!t]
    \centerline{\includegraphics[width=\textwidth]{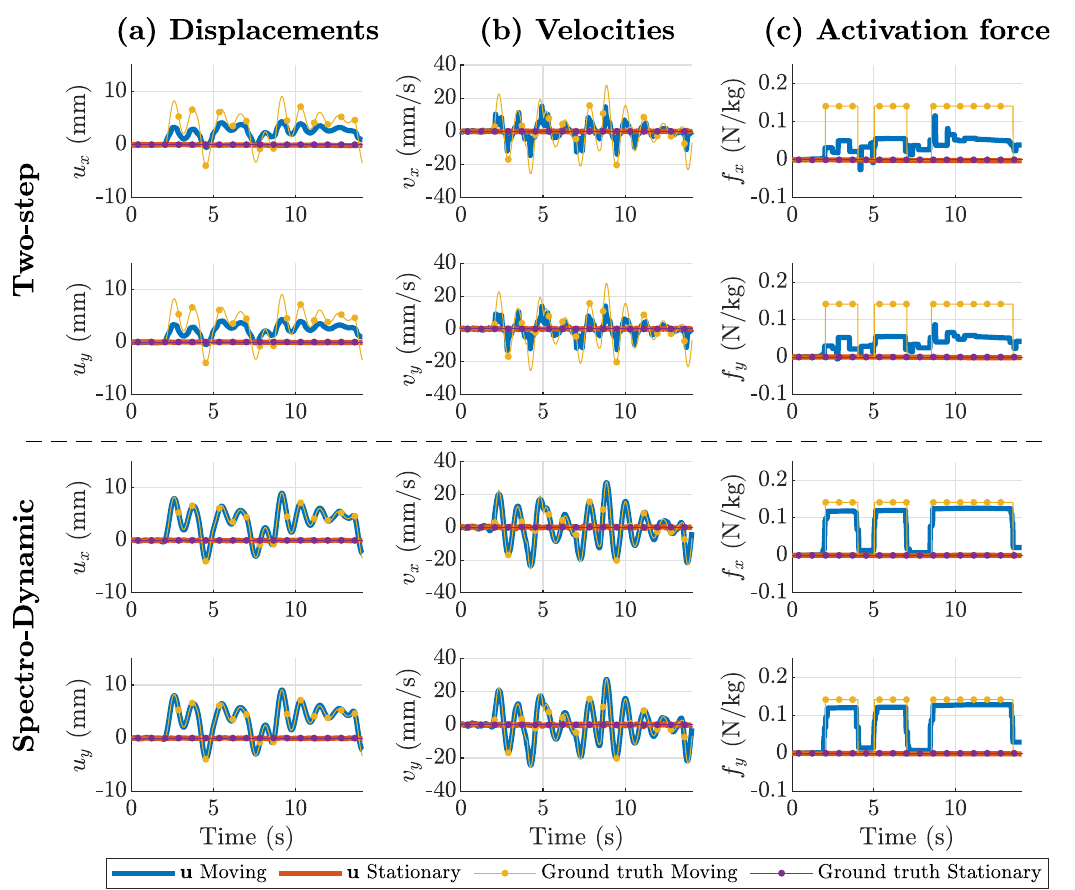}}
    \caption{Estimated (a) displacements, (b) velocities, and (c) activation forces in both directions resulting from the two-step reconstruction (top) and Spectro-Dynamic reconstruction (bottom) for the experiment with a discontinuous activation function and $\orient = 45\degree{}$. The velocities were calculated from the displacements using finite differences. The motion of both the moving and the stationary compartment was estimated, but only the moving compartment moved along one direction. Since $\orient=45\degree{}$, this motion was separated over the $x$- and $y$-directions.}
    \label{sfig:dyn-discont-45}
\end{figure*}

\begin{figure*}[!t]
    \centerline{\includegraphics[width=\textwidth]{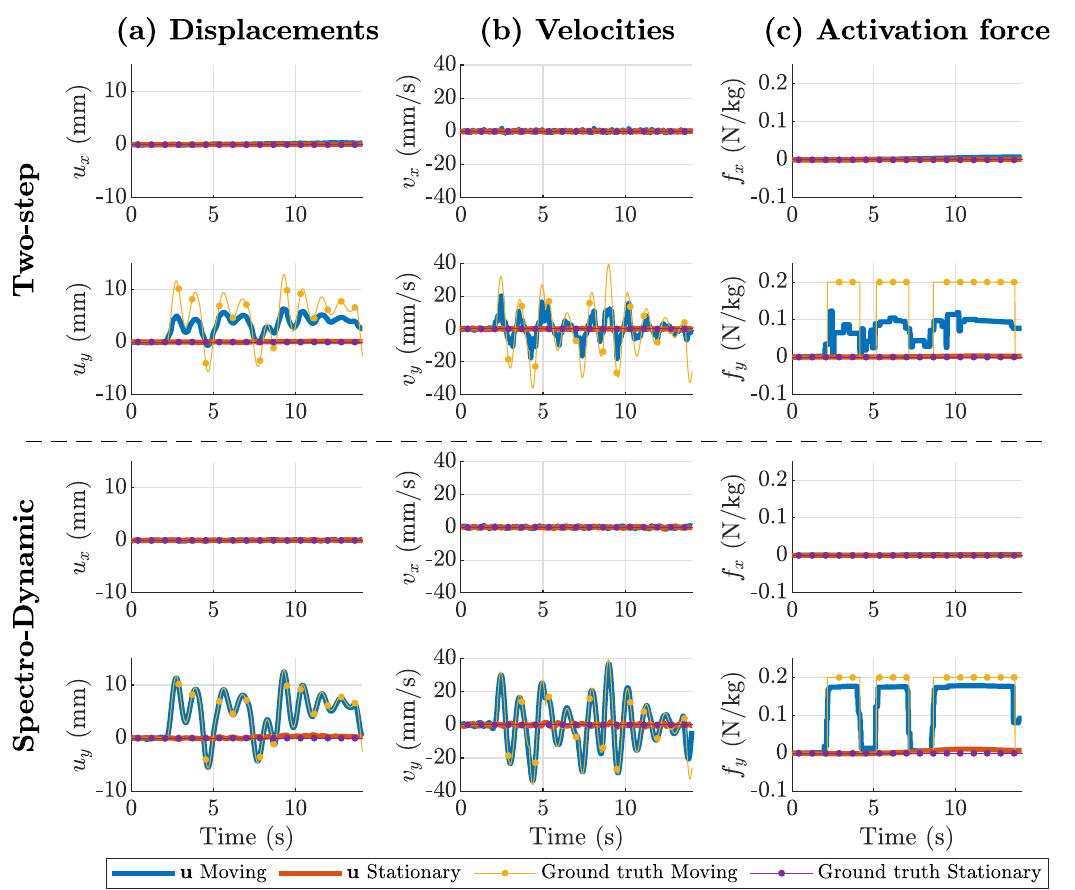}}
    \caption{Estimated (a) displacements, (b) velocities, and (c) activation forces in both directions resulting from the two-step reconstruction (top) and Spectro-Dynamic reconstruction (bottom) for the experiment with a discontinuous activation function and $\orient = 90\degree{}$. The velocities were calculated from the displacements using finite differences. The motion of both the moving and the stationary compartment was estimated, but only the moving compartment moved along one direction ($y$).}
    \label{sfig:dyn-discont-90}
    \label{sfig:dyn-last}
\end{figure*}

\makeatletter
\makeatletter \renewcommand{\fnum@figure} {Video~S\thefigure} 
\makeatother
\setcounter{figure}{0}

\begin{figure*}[!t]
    \centerline{\includegraphics[width=\textwidth]{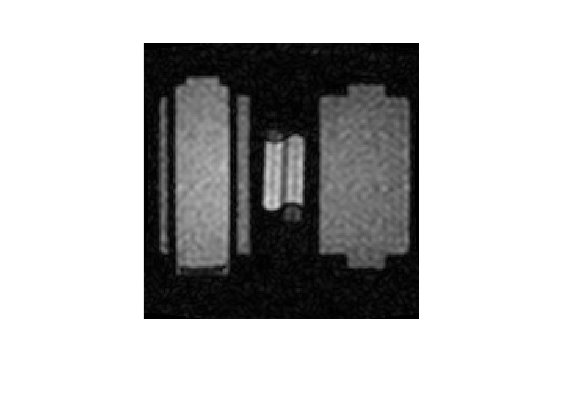}}
    \caption{Video showing the time-resolved image data. Each frame is obtained by performing a Fast Fourier Transform (FFT) on one time instance of the estimated time-resolved k-space data $\fsd$, followed by a geometry correction to account for gradient nonlinearities. The video can be viewed online, this document only shows a single frame.}
    \label{svid:recon-cont-0}
\end{figure*}